\documentclass[prl,twocolumn,showpacs]{revtex4-1}

\usepackage{amsmath}
\usepackage{amssymb}
\usepackage{amsthm}

\usepackage{graphicx}
\usepackage{graphics}
\usepackage{bbold,bm}
\usepackage{hyperref}
\usepackage{epsfig}
\usepackage{dcolumn}

\begin{document}

\title{Effective Theory of Floquet Topological Transitions}

\author{Arijit Kundu}
\affiliation{Department of Physics, Indiana University, Bloomington, IN 47405}
\author{H.A. Fertig}
\affiliation{Department of Physics, Indiana University, Bloomington, IN 47405}
\author{Babak Seradjeh}
\affiliation{Department of Physics, Indiana University, Bloomington, IN 47405}

%\date{\today}

\begin{abstract}
We develop a theory of topological transitions in a Floquet topological insulator, using graphene irradiated by circularly polarized light as a concrete realization.
We demonstrate that a hallmark signature of such transitions in a static system, i.e. metallic bulk transport with conductivity of order $e^2/h$, is substantially suppressed at some Floquet topological transitions in the clean system. We determine the conditions for this suppression analytically and confirm our results in numerical simulations. Remarkably, introducing disorder dramatically enhances this transport by several orders of magnitude.
\end{abstract}

 \maketitle

%---- Introduction ----%
\emph{Introduction}.---%
Topological insulators exhibit a variety of phenomena usually
associated with quantum behaviors of electrons in strong magnetic
fields, even though no such fields are present~\cite{kane_2005,hasan_2010,qi_2011}. A characteristic aspect of these phenomena is the so-called bulk-boundary correspondence due to the appearance of robust states at the edges of the system, whose precise structure is determined by the topology of the bulk state. In particular, edge transport in a four-terminal geometry~\cite{kitagawa_2011} can reveal the change in bulk topology by yielding sharply different results depending on the number and chirality of the edge states in different gapped states. At a topological phase transition the bulk band gap closes and the low-energy dynamics is controlled by a gapless state. Therefore, bulk metallic behavior also marks the change in the topological state.

Much of the search for these remarkable systems has focused on materials
with strong spin-orbit coupling.  Recently, it has been appreciated that periodically driven systems can exhibit a rich topological phase diagram as the frequency, amplitude, and other properties of the drive are tuned~\cite{oka_2009,gu_2011,lindner_2011,JiaKitAli11a,KunSer13a}. In this way, a much
simpler system can support topological states even if the static system 
does not \cite{lindner_2011,dora_2012}. A prime candidate for realizing such a Floquet topological insulator is graphene~\cite{Castro_Neto_RMP,peres_2010,dassarma_2011} in the
presence of circularly polarized
light~\cite{oka_2009,gu_2011,roslyak_2011,kitagawa_2010,kitagawa_2011,calvo_2011,delplace_2013}
or other temporally periodic potentials \cite{lopez_2013,iadecola_2013}.
Experimental evidence for a similar Floquet state on the surface of Bi$_2$Se$_3$ has been reported in Ref.~\onlinecite{wang_2013}. The resulting topological state is intrinsically dynamical and the correspondence between bulk and edge state structure is more subtle than in the static system. Thus, a natural question arises as to the signature of the critical state that is realized at the transition between Floquet topological states.

In this paper we show that while the low-energy critical dynamics in irradiated graphene is controlled by a massless Dirac Hamiltonian, the metallic transport across the system can be {\it substantially} suppressed, with conductivity much smaller than that of static systems,
which is of order $e^2/h$ ~\cite{Tworzydlo_2006}. Hence, some topological transitions in Floquet topological insulators may be concealed in the 
bulk system while the edge signatures are intact.
Remarkably and counterintuitively, disorder can increase conductivity by several orders of magnitude and thus unmask the topological transition.
We demonstrate that the reason for this intricate behavior lies in an important contrast between Floquet bands in a time-dependent
two-dimensional system and energy bands of a static one.  Specifically, the topological transitions in the former case are most naturally described by
a $2+1$ dimensional theory \cite{leon_2013}, in which the extra (time) dimension is effectively compactified.  The extra degree of freedom, present in the Floquet
system but absent in static leads, creates a mismatch between the two
with sometimes dramatic consequences for transport.  Our theory also naturally explains why the phase diagram of topological states in such systems
is, as we find, remarkably rich.

%--------- Fig 1. ----------%
\begin{figure}
  \includegraphics[width=3in]{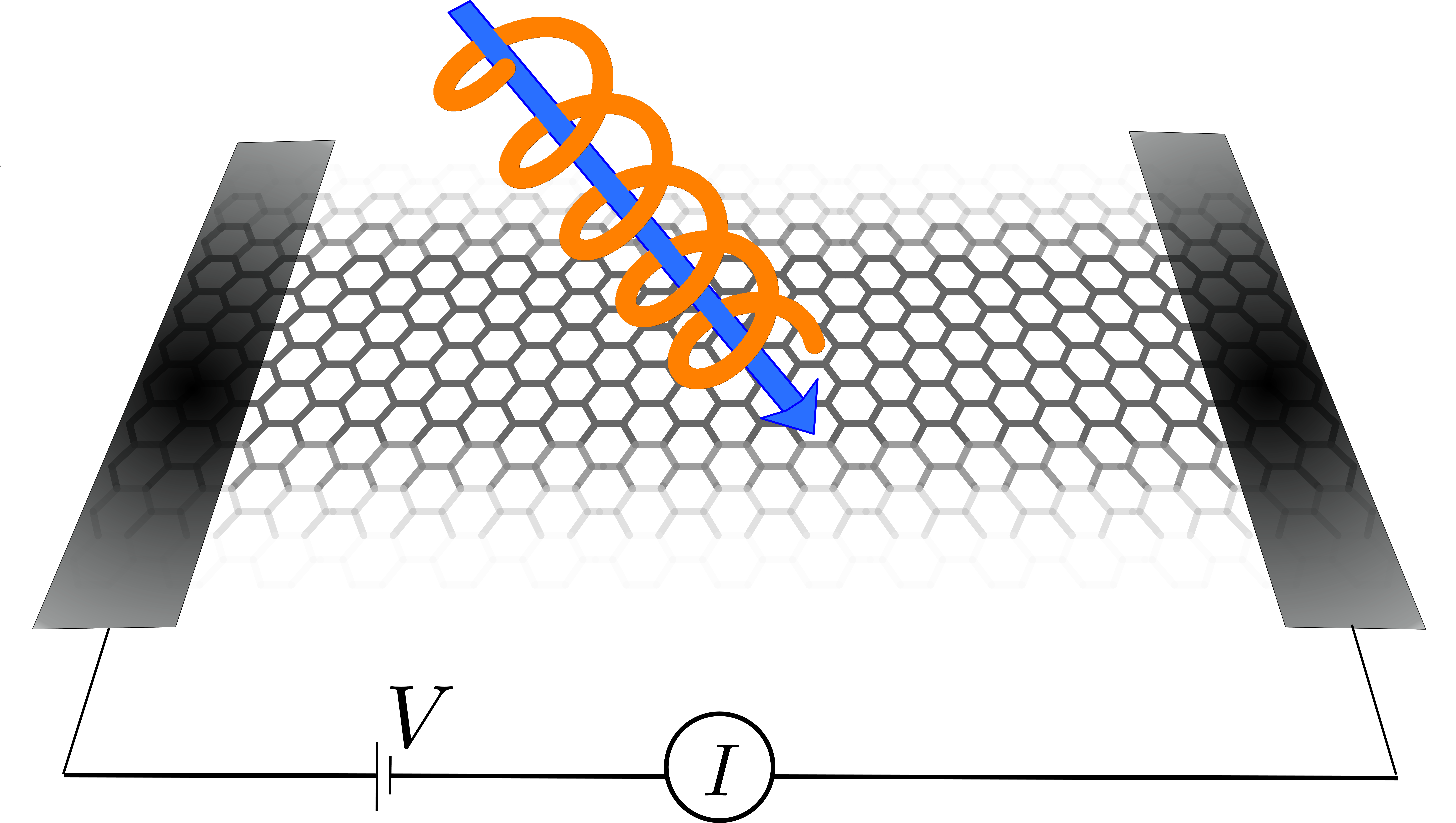}
  \caption{(color online)
  Setup and transport geometry.}
   \label{fig:setup}
\end{figure}

%---- Model and Topological Transitions ----%
\emph{Model and Bulk Topological Transitions}.---%
The geometry we consider is illustrated in Fig.~\ref{fig:setup}.
Circularly polarized light of frequency $\Omega$ and electric
field strength $\Omega A_0/c$, with $c$ the speed of light, impinges
on a large graphene sheet.  The field is assumed to be uniform in
the plane so that the electronic states may be characterized by
a two-dimensional wavevector {\bf k} of a band $b$ within the
hexagonal Brillouin zone (BZ).  Because the vector potential is
periodic, the time-dependent Schr\"odinger equation represents
a Floquet problem \cite{rahzavy_2003}, for which wavefunctions $\psi$,
two-component spinors encoding the wavefunction amplitudes on each of
the sublattices of the honeycomb lattice,
can be represented in the form
$
\psi_{{\bf k} , b}({\bf r},t)= u_{b}({\bf k},t)
e^{i{\bf k} \cdot {\bf r} - i\varepsilon_{b}({\bf k}) t},
$
with $u_{b}({\bf k},t+2\pi/\Omega)=u_{b}({\bf k} , t)$
a periodic function.  This obeys the (2+1)-dimensional eigenvalue equation,
$
H_F({\bf k},t)  u_{b}({\bf k},t) = \varepsilon_{b}({\bf k}) {u}_{b}({\bf k},t),
$
with Floquet Hamiltonian of the form ($\hbar=1$)
%%%%%
\begin{equation}
H_F({\bf k},t)=
\left(
\begin{array}{cc}
-i\partial_t & -\gamma Z({\bf k},t)  \\
-\gamma Z^*({\bf k},t) & -i\partial_t
\end{array}
\right).
\label{HF}
\end{equation}
%%%%%
Here, $\gamma$ is the hopping parameter, $Z({\bf k},t)=\sum_{n=1}^3
e^{i[{\bf k}+{ e \over c} {\bf A}(t)] \cdot {\bf a}_n }$ is the nearest-neighbor tight-binding factor where
${\bf a}_n=a_0(\cos\theta_n,\sin\theta_n)$, with $\theta_n=(2n-1)\pi/3$, %$\theta_1=\pi$, $\theta_2=\pi/3$, $\theta_3=-\pi/3$
are the three nearest neighbor vectors of the honeycomb lattice for graphene
(lattice parameter $a_0=0.142$nm) and ${\bf A}(t)=A_0(\cos\Omega t,\sin\Omega t)$
is the vector potential of a rotating electric field.
The quasienergies $\varepsilon_{b}(\bf k)$
can be chosen in the ``Floquet zone'' (FZ)
$-\Omega/2 < \varepsilon_{b}({\bf k}) \le \Omega/2$
by multiplying $ u_{b}({\bf k},t)$ by $e^{in\Omega t}$ with
integer $n$ appropriately selected.  While solutions related to one another
in this way represent the {same} solution to the time-dependent
Schr\"odinger equation, they are {\it different} eigenstates of $H_F$.
The full vector space of these states is needed to allow matching
of wavefunctions to external states, for example to a lead where
current may be injected or removed.  Thus, to deal with such systems
one must treat the time degree of freedom as a genuine extra dimension,
compactified via the periodic (temporal) boundary condition on ${u}_{b}$.

%--------- Fig 2. ----------%
\begin{figure}
  \includegraphics[width=3in]{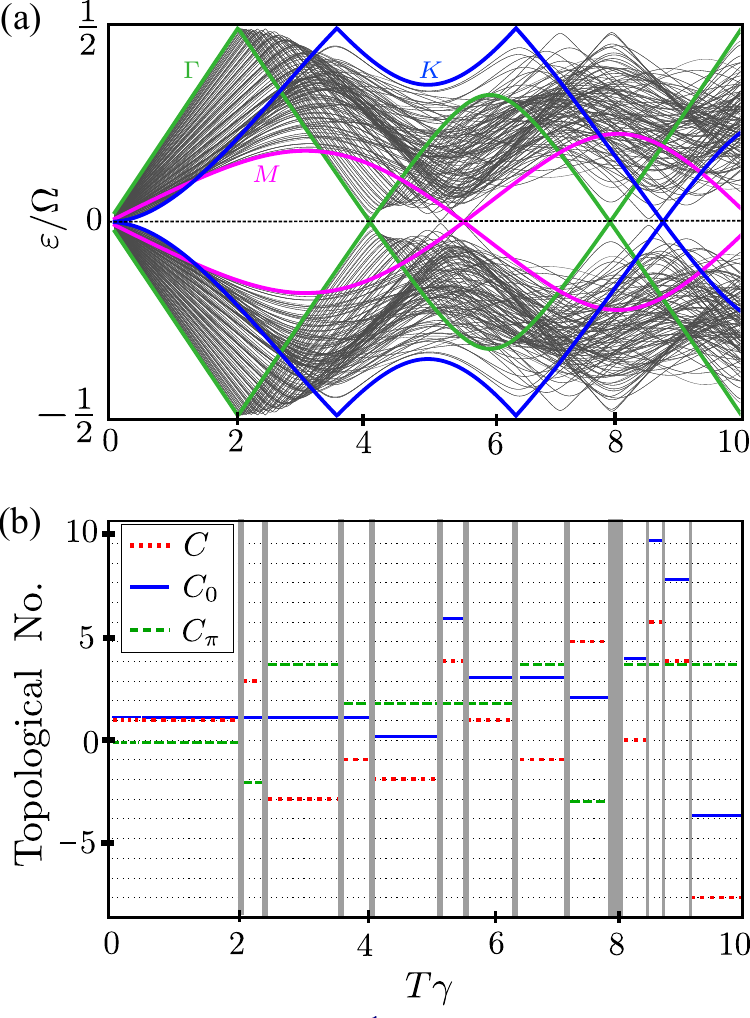}
  \caption{(color online)
  Bulk topological transitions and invariants. (a) Bulk quasienergies vs. period showing several topological transitions where
  quasienergies cross zero and/or $\Omega/2$. The solid (green, blue and magenta) lines trace the evolution of the $\Gamma$, $K$ and $M$ points. (b) Bulk Chern numbers $C_0$ and $C_\pi$
  and the total Chern number $C=C_0-C_\pi$ vs. period. The shaded gray areas are the topological transitions where the numerical error in computing the Chern numbers is large.}
   \label{fig:bulkT}
\end{figure}

Fig.~\ref {fig:bulkT}(a) illustrates
numerically generated values of $\varepsilon_{b}({\bf k})$
associated with $H_F$ vs. period $T=2\pi/\Omega$. Several transitions are observed where
quasienergy gaps close around zero or $\Omega/2$.
The first topological transition for $T>0$ occurs when the gap at the FZ boundary $\Omega/2$
closes at the $\Gamma$ point (${\bf k}=0$).
Further topological transitions occur as $T$ increases, and may proceed in several
ways, always involving a gap closing.  In most cases the closings involve
the formation of a Dirac point either at zero quasienergy or at the FZ boundary.  These may occur at BZ high-symmetry points,
$\Gamma$, $K$, or $M$, yielding, respectively one, two, or three distinct
closing points. As explained in the next section, the gap closing at
$\Gamma$ can sometimes be quadratic.  We also observe gap closings away from high-symmetry points, in which case we observe six distinct Dirac points at the transition.

These gap closings are accompanied by changes in the topology of the bulk band structure characterized by its Chern numbers. The Chern number $C$ of each Floquet band is calculated by integrating its Berry's curvature. Because each Dirac point induces an exchange of Chern number $\pm 1$ between
bands, and a quadratic touching induces an exchange of $\pm 2$ across the transition,
all our transitions involve Chern number changes of magnitude $|\Delta C|=1,2,3$ or 6~\cite{murthy_2012}.
Bulk-boundary correspondence in a static system relates the Chern number to the total number of edge states.
However, in a Floquet system, the two gap closings at zero quasienergy and at FZ boundary
are not equivalent: when edges are present in the system, changes in
the number of edge states are induced around the quasienergy
in which a gap closing occurs.  This means there may be
topologically distinguishable states with the same Chern number.
Specifically, $C=C_0-C_\pi$, where $C_0$ and $C_\pi$ are topological invariants that relate, respectively, to edge states traversing the gaps 
at zero quasienergy and the FZ boundary~\cite{rudner_2013}.
Our model allows numerical computation of $C_0$ and $C_\pi$,
fully specifying the bulk topology and determining where edge states
lie in the quasienergy spectrum~\cite{suppl}.  This
results in a particularly rich phase diagram for the system as its
parameters are varied.
Fig.~\ref{fig:bulkT}(b) shows the associated Chern numbers calculated numerically, clearly identifying the topological nature of the transitions. We have also explicitly checked that the edge state spectrum in a finite ribbon agrees with the Chern numbers. Our exploration of this phase diagram is
detailed in the Supplemental Material~\cite{suppl}.

\emph{Topological Transitions in a Floquet Tight-Binding System}.---%
We next develop a framework to describe these topological
transitions and their transport signatures, in a way that is
generalizable to any tight-binding system.  We begin with the
Floquet Hamiltonian, which for the graphene system is given
by Eq.~(\ref{HF}).
In general
the quantity $Z({\bf k},t)$
may be expanded in a Fourier decomposition,
$Z({\bf k},t)=\sum_m Z_m({\bf k}) e^{-im\Omega t}$. For concreteness we consider the transitions at the $\Gamma$ and $K$ points.

Precisely at the
$\Gamma$ point $Z$ has a simple form: for any integer $m$,
$Z_{3m+1}({\Gamma})=Z_{3m+2}({\Gamma})=0$, and
$Z_{3m}(\Gamma)=\pm3 J_{3m}(\alpha)$, where $J_{3m}$ is a Bessel
function,  $\alpha= eA_0a_0/c$, and
the upper (lower) sign applies for
$m$ even (odd).  Because $Z_{m}(\Gamma) \ne 0$ only
for $m$ divisible by 3, the $\Gamma$ point has a remarkable property:
at this point, $H_F$ has a period only 1/3 that of other $k$-points.
This effect is an entwining of the temporal and spatial symmetries,
and it allows different states with quasienergy $\pm \Omega/2$
to be degenerate for appropriately chosen parameters.  When this occurs,
the system is topologically critical, as we show below.
To the lowest order in $\alpha$, we may keep only $Z_0(\Gamma)$ and
$H_F(\Gamma)$ becomes nearly time-independent.  Its eigenvalues
are approximately given by $\varepsilon_{\pm}^{(m)}(\Gamma)=m\Omega \pm 3\gamma J_0(\alpha)$,
and the associated eigenvectors are just
%%%%%
$$
{u}^{(m)}_{\pm}(\Gamma,t)={1 \over \sqrt{2T}} e^{im\Omega t} {1 \choose \pm1}.
$$
%%%%%
The first crossing is that of the $m=\pm 1$ level with the $m=0$ level at the FZ boundary,
followed by the crossing between $m=1$ and $m=-1$ at zero quasienergy.

\begin{figure}
  \includegraphics[width=3in]{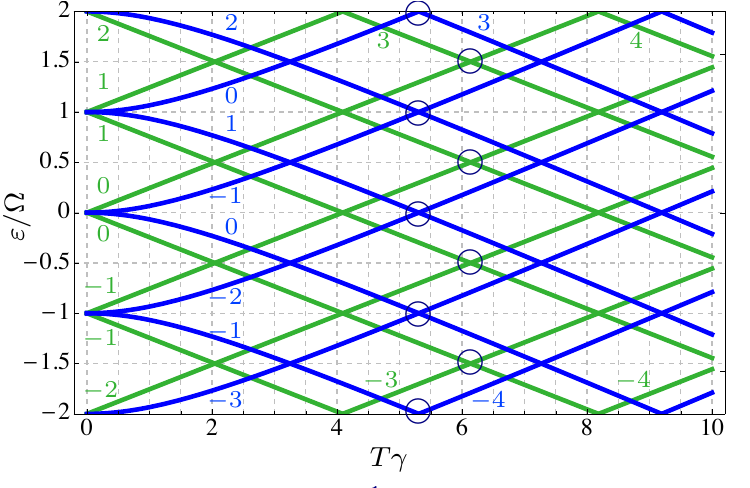}
  \caption{(color online)
  Quasienergy spectrum at $\Gamma$ (green) and $K$ (blue) points obtained from the lowest perturbation theory. The Fourier index $m$ is shown.
  Circles mark the avoided crossings between branches where $m$ differs by $\pm3$.
  }
  \label{fig:theory}
\end{figure}

At the $K$ point, an analogous calculation yields $Z_{3m}(K)=Z_{3m+1}(K)=0$, $Z_{3m+2}(K)= (-)3 J_{3m+2}(\alpha)$ for $m$ even (odd). The largest contribution now comes from $Z_{-1}(K)$ 
and the eigenvalues are approximately $\varepsilon_\pm^{(m)}(K)=(m+\frac12)\Omega\pm\sqrt{(\Omega/2)^2+[3\gamma J_1(\alpha)]^2}$. The eigenvectors have a form
%%%%%
$$
{u}^{(m)}_{\pm}(K,t)={1 \over \sqrt{2T}} e^{im\Omega t} {e^{i\Omega t} u_A \choose \pm u_B},
$$
%%%%%
with time-independent $u_A$ and $u_B$.
The first transition appears at the FZ boundary between $m=-1$ and $m=1$ at $\varepsilon=\Omega/2$ and $m=-2$ and $m=0$ at $\varepsilon=-\Omega/2$. 
Fig.~\ref{fig:theory} illustrates the eigenvalues as a function of $T$ for $\alpha=1.5$.
The periods agree well with critical values we find numerically  [Fig.~\ref{fig:bulkT}(a)].

To infer the order of a transition, we project the full Floquet Hamiltonian
into a two-dimensional space for each value of ${\bf k}$ in the vicinity of
the BZ point where the transition occurs using the states of the form ${u}^{(m)}_{\pm}
e^{i{\bf k} \cdot {\bf r}}$, in precise analogy with ${\bf k} \cdot {\bf p}$
perturbation theory.  At the $\Gamma$ point near the transition at zero quasienergy, the
 $m= \pm 1$ states are degenerate and their eigenvectors are used to  construct basis states in the vicinity of the $\Gamma$ point.
The resulting
projected Hamiltonian is $\overline{H}_F={\bf h}({\bf k})\cdot\boldsymbol{\sigma}$, where $\boldsymbol\sigma$ is the vector of Pauli matrices and
%%%%%
\begin{eqnarray}\label{crit_h_2}
h_{x}-ih_{y} &=& -i{{3\gamma} \over 2} J_2(\alpha)(k_x+ik_y)a_0, \\
h_{z} &=& -\Omega+3\gamma J_0(\alpha)(1-{1 \over 4}k^2a_0^2),
\end{eqnarray}
%%%%%
corresponding to $|\Delta C| = 1$ around
$\Omega=3\gamma J_0(\alpha)$, in good agreement with our numerics.
One may perform the same kind of analysis near the first transition at the FZ boundary. To second order in ${\bf k}$, this produces the projected
Hamiltonian $\overline{H}_F={\Omega \over 2}+{\bf h}({\bf k})\cdot\boldsymbol{\sigma}$, with
$h_{x}-ih_{y} \sim (k_x-ik_y)^2$. (See Supplemental Material for details~\cite{suppl}.)
One may easily confirm that there is a $|\Delta C| = 2$ exchange
between the two bands that diagonalize this Hamiltonian when
$\Omega/2 = 3\gamma J_0(\alpha)$.  This is precisely the change we observe
in our numerics.% around $T\gamma\approx2.1$ for $\alpha=1.5$.

%--------- Fig 3. ----------%

Of course, not all the gap closings occur at the $\Gamma$ and $K$ points. 
In principle the method we have developed can be employed to describe
the region around a gap closing in terms of just two (time-dependent) states,
from which the order of the transition and its Fourier structure can be inferred.  Furthermore,
there is nothing fundamental about our approach limiting it to graphene:
it can be employed for general systems in which Floquet band-gap closings occur.

\emph{Transport Properties and Disorder}.---%
As discussed above, bands of different $m$ values, representing different 
temporal subbands of the $2+1$-dimensional problem, get woven into the quasienergy 
bands in the FZ as topological transitions occur.
The resulting states have a
 complicated time dependence, sometimes allowing
only very weak couplings to states governed by a 
static Hamiltonian (for example, those
of a time-independent lead),
since the latter have much simpler time dependence.
In such cases, at a transition---where one expects 
conductivity of order $e^2/h$ in a two-terminal geometry~\cite{Tworzydlo_2006} 
when the Fermi energy is at a Dirac point---the conductivity will be suppressed. Moreover, as we show below,
disorder can spoil the entwining of spatial and temporal symmetries that causes this suppression.

%--------- Fig 4. ----------%
\begin{figure}
  \includegraphics[width=0.47\textwidth]{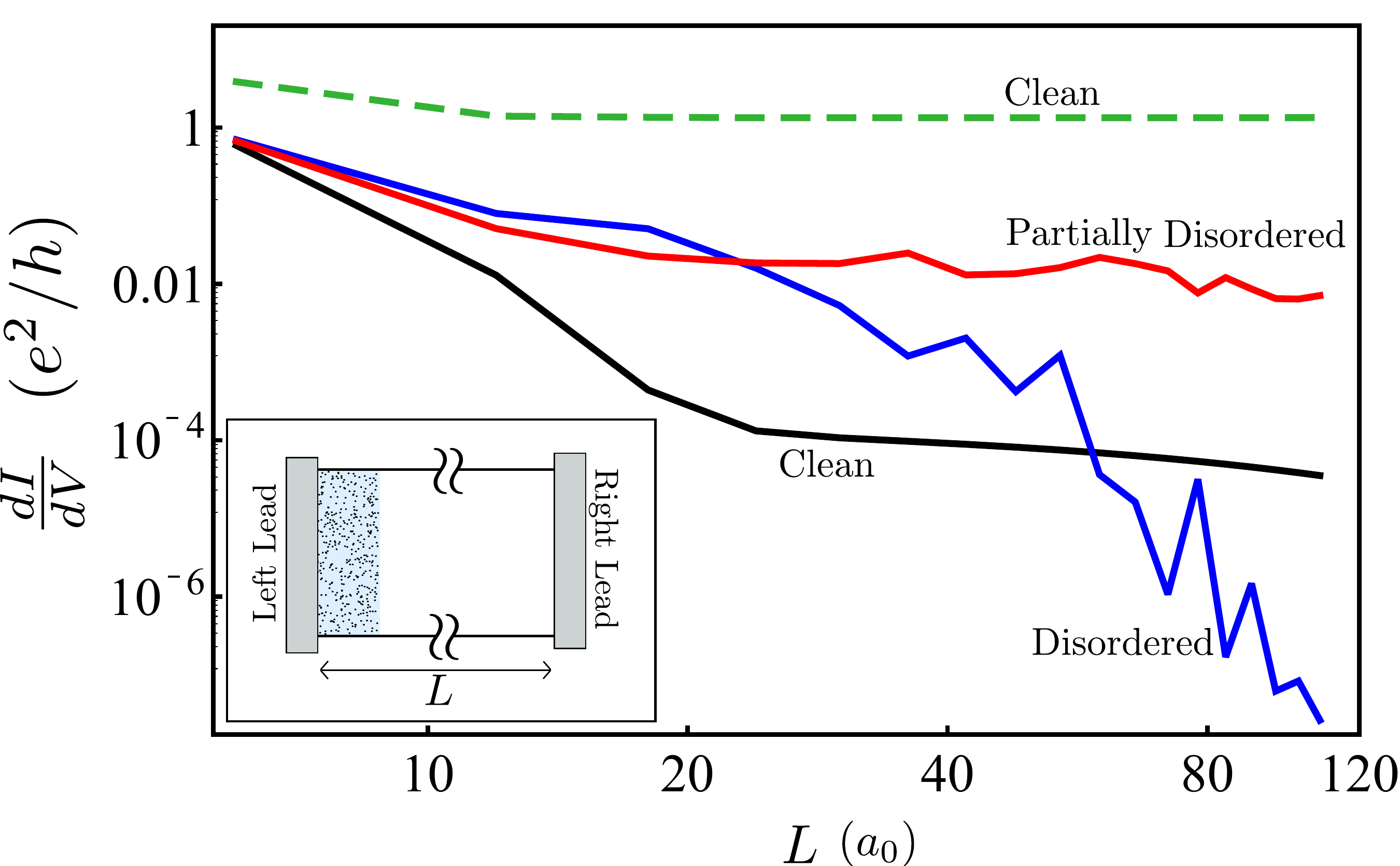}
  \caption{(color online)
  Bulk transport. Differential conductance vs. distance between the leads in an undoped ribbon at the $\Gamma$ point (solid lines) and $K$ point (dashed green line) transitions at, respectively, $T\gamma \approx 4.1$ and $T\gamma\approx3.2$ when $\alpha=1.5$. Three different situations are shown for the $\Gamma$ point: a clean system (black), a partially
  disordered one (red) where on-site disorder is present only over a length 9$a_0$ in the contact region (see inset), and one in which there is disorder throughout the ribbon (blue).  The disorder strength in both cases is 0.15$\gamma$.
  }
  \label{fig:transport}
\end{figure}

An example of these phenomena is illustrated
in Fig.~\ref{fig:transport}, where we present 
differential conductance results in the system
with periodic boundary conditions along its width as a function of length $L$ for $\alpha=1.5$~\footnote{
In a driven system and for a generic geometry, broken time reversal symmetry will allow
charge pumping between two leads even in the absence of a bias.
One may nevertheless probe the system by considering the
{\it change} in current due to interlead bias, i.e., by measuring the
differential conductance.  Moreover, clean systems with smooth parallel boundaries
can have sufficient symmetry to suppress charge pumping at zero bias~\cite{gu_2011}.}.
Details of the calculation scheme are discussed
in the Supplemental Material~\cite{suppl}. In the clean system, the transition at the $\Gamma$ point for
$T\gamma\approx4.1$ presents a conductance that drops
precipitously with $L$---similar to what one would expect if the spectrum
were gapped---before leveling off at $\sim 10^{-4}e^2/h$, yielding behavior
roughly like that of a very poor metal.  By contrast, the differential conductance at the $K$ point transition for $T\gamma\approx3.2$
has a scale and $L$ dependence similar to that of a static system.

The surprising behavior of the $T\gamma\approx4.1$ transition reflects
the 2+1 dimensional nature of the Floquet system.  As illustrated
in Fig.~\ref{fig:theory}, the transition involves a degeneracy at zero quasienergy at the $\Gamma$ point for temporal subbands
$m=\pm 1$, so that these states have time dependence $\sim e^{\pm i\Omega t}$.
Small corrections to this exist at high order in $\alpha$, but the symmetry
of the $\Gamma$ point allows only subbands with time dependence
$e^{i(3n \pm 1)\Omega t}$ to enter the exact eigenstate of $H_F$.
By contrast, states in the lead at the Fermi energy $E_F=0$ have {\it no}
time dependence.  Since wavefunctions must match at a junction
between a lead and the ``scattering'' region in both space and time,
only the temporal $m=0$ state at the $\Gamma$ point matches onto the relevant lead state.  Because this does not occur at zero quasienergy,
the corresponding state
is evanescent, leading to the dramatic falloff with $L$.  Away from the
$\Gamma$ point, symmetry does allow some admixture of an $m=0$ temporal
state, and a corresponding differential conductance that falls off
approximately as a power law with $L$ rather than exponentially.  Since
this coupling vanishes precisely where the gap vanishes, the net
result is an anomalously small conductance at large $L$.  By contrast, gap closings where a time-independent subband participates yield far
larger conductances.  The first $K$ point transition at $T\gamma\approx3.2$ provides an example of this, in agreement with our numerical results
 in Fig.~\ref{fig:transport}.

A remarkable consequence of this structure is that the differential conductance may be
greatly \emph{enhanced} by disorder, which spoils the entwined spatial-temporal symmetry at the $\Gamma$ point,
and allows the $m=0$ subband to be admixed into states with quasienergy
near zero.  As illustrated in Fig.~\ref{fig:transport} for the $T\gamma\approx4.1$ transition, this is particularly effective
when  disorder is concentrated near the junction to the lead (see inset).
Disorder throughout the ribbon yields very similar behavior at small
$L$, but suppresses the conductance at large $L$, a phenomenon we
associate with localization of the wavefunctions.

\emph{Discussion}.---%
In this study we have focused on results for ribbons with periodic boundary
conditions along their width in order to understand bulk transport. The system with open boundary conditions also supports edge
states whose structure depends on the topological indices of the state.
Our numerics confirms this behavior, as well as their contribution to
transport.  It should be noted however that in the infinite width limit,
edge state transport is always negligible compared to bulk transport
for metallic states
in a ribbon geometry.  Away from a transition point, multi-terminal geometries allow one to directly
probe edge states and 
determine different topological state of the system~\cite{kitagawa_2011},
even as transitions between them may be masked in two-terminal experiments.

Disorder enhanced conductance is one of the very unusual behaviors of a Floquet
topological insulator.  It is a direct consequence of multiple subbands in the time
dimension of the problem, which endows the system with a rich set of
topological phases and transport properties with no analog in static, two-dimensional
topological insulators.

This work was supported in part by the NSF through
Grant No. DMR-1005035 and by the US-Israel
Binational Science Foundation.  Further funding was provided by the College of Arts and Science and the
Offices of the Vice President for Research and the
the Vice Provost for Research at Indiana University
through the Faculty Research Support Program.
The authors are grateful to Takashi Oka for a number of helpful discussions.

\renewcommand{\thefigure}{S\arabic{figure}}
\setcounter{figure}{0}
\renewcommand{\theequation}{S\arabic{equation}}
\setcounter{equation}{0}

\section{Supplemental Material}

\subsection{Topological invariants}
Here we briefly define the topological invariants associated with a two-dimensional time periodic system with Bloch Hamiltonian $H(\mathbf{k},t) = H(\mathbf{k}, t+T )$, characterized by the time evolution operator $$U(\mathbf{k},t) = \mathcal{T} \exp\left(-i\int_0^t dt' H(\mathbf{k},t') \right).$$
If for a system $U(\mathbf{k},T) 
= U(\mathbf{k},0) = \mathbf{1}$ then $U$ defines a map from the 3-torus $\mathbb{T}^3$ to the unitary group $\mathbb{U}(N)$, where $N$ is the dimension of $U$. These maps are known to be classified by an integer winding number 
\begin{align}\label{eq:winding}
 W[U] =& \frac{1}{8\pi^2}\int dt d{\bf k} \nonumber \\
& \times \mathrm{Tr}\left(U^{-1}\partial_tU[U^{-1}\partial_{k_x}U, U^{-1}\partial_{k_y}U] \right).
 \end{align}
But, in general $U(\mathbf{k},T) \neq U(\mathbf{k},0)$ and one needs to construct a trivial time evolution operator $U_{\varepsilon}(\mathbf{k},t) = \mathbf{1}$ such that $U_{\varepsilon}$ can be interpolated smoothly to the original time evolution operator without closing a gap around the quasienergy $\varepsilon$.

We follow the prescription provided in Ref. \onlinecite{rudner2013} to compute the `modified' time evolution operator as
\begin{align}
 U_{\varepsilon}(\mathbf{k},t) &= \left\{ \begin{array}{ll}
                                          U(\mathbf{k},2t) & ~~\text{if}~0\leq t\leq T/2 \\
                                          V_{\varepsilon}(\mathbf{k},2T-2t) & ~~\text{if}~T/2\leq t\leq T,
                                         \end{array}\right.
\end{align}
where
\begin{align}
 V_{\varepsilon}(\mathbf{k},t) &= \exp\Big(-iH_{\text{eff}}(\mathbf{k})t\Big), \quad H_{\text{eff}}(\mathbf{k}) = \frac{i}{T} \log U(\mathbf{k},T).
\end{align}
Here, one chooses the branch-cut according to
\begin{align}
 \log e^{-i\varepsilon T +i0^-} &= -i\varepsilon T \nonumber \\
 \log e^{-i\varepsilon T +i0^+} &= -i\varepsilon T - 2\pi i. \nonumber
\end{align}

We define the topological invariants $C_0$ and $C_{\pi}$ as following 
\begin{align}\label{eq:inv}
 C_0 = W[U_0], \quad C_{\pi} = W[U_{\pi/T}].
\end{align}
Finally, one can show the relation between the winding number and the Chern number is~\cite{rudner2013} $$ W[U_{\varepsilon}] - W[U_{\varepsilon'}] = C_{\varepsilon',\varepsilon}, $$
where $C_{\varepsilon',\varepsilon}$ denotes the sum of Chern numbers of all Floquet band(s) that lie in between $\varepsilon'$ and $\varepsilon$. Considering all Floquet band(s) below zero quasienergy filled, we define $C = C_{0,-\pi/T} = - C_{\pi/T,0}$, which gives us the relation $$C_0 - C_{\pi} = C.$$
We must point out that the numerical computation of the integral in Eq.~(\ref{eq:winding}) does not necessarily give an integer as opposed to the computation of Chern number~\cite{Fukui2005}. But the integration asymptotically converges to an integer as one increases the density of sample points.

\begin{figure*}[tb]
  \includegraphics[width=6in]{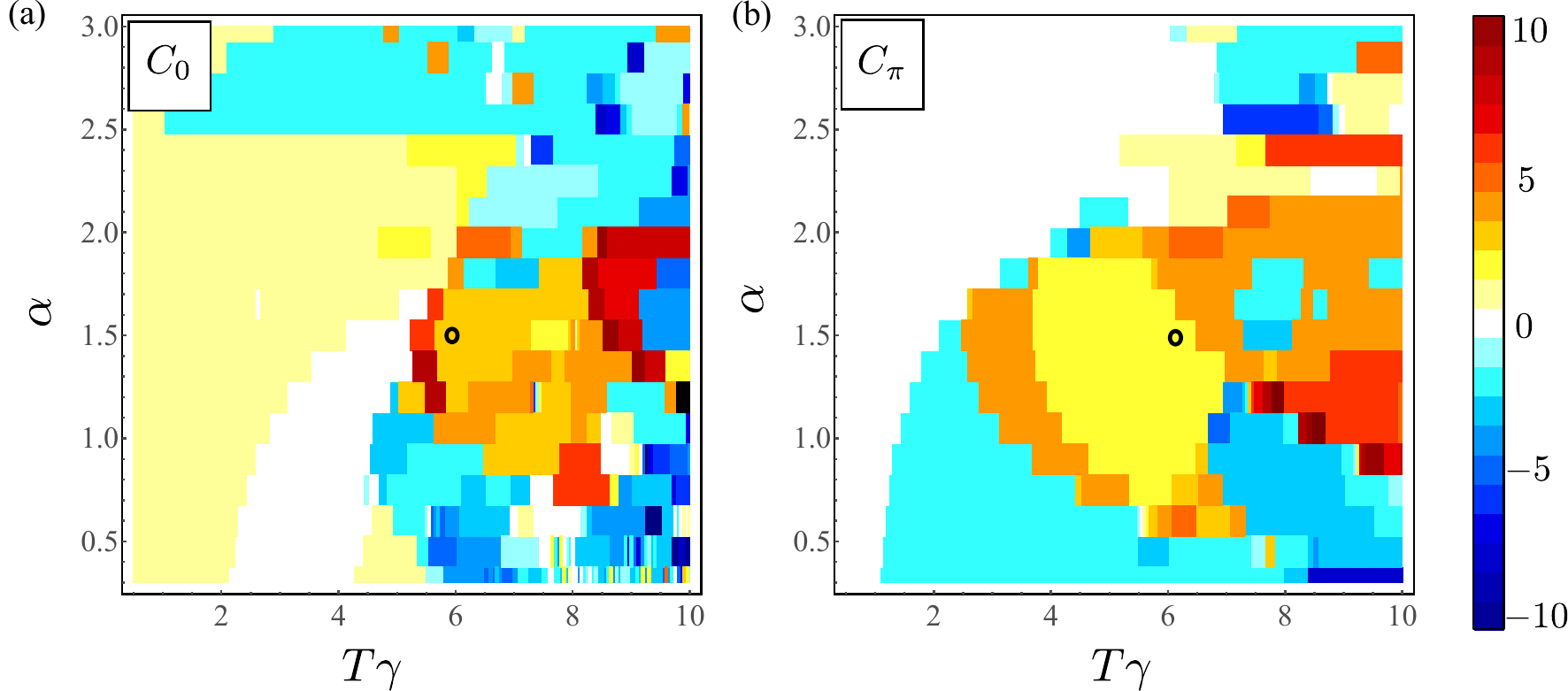}
  \caption{Topological invariants $C_0$ (a) and $C_{\pi}$ (b) as a function of $\alpha$ and $T\gamma$. The total Chern number of the Floquet bands below zero quasienergy is $C = C_0 - C_{\pi}$. The circles denote the value of the parameters used in Fig.~\ref{fig:ribbon}.} 
    \label{fig:phase}
\end{figure*}

\begin{figure}
  \includegraphics[width=3in]{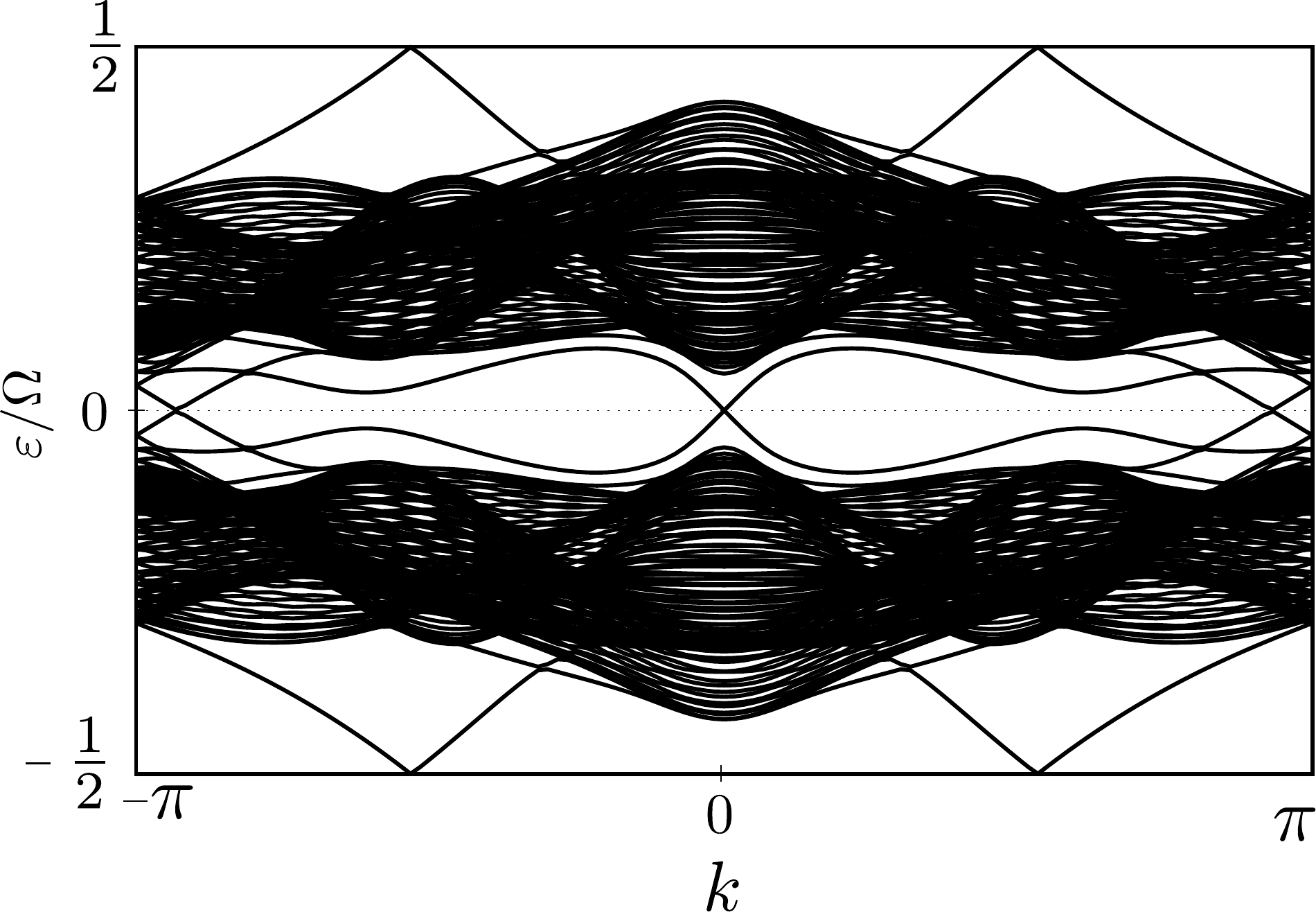}
  \caption{Quasienergies of a ribbon with an edge along the armchair direction for $\alpha=1.5$ and $T\gamma = 6$. With these parameters we have $C_0 = 3$ and $C_{\pi} = 2$. Precisely, 3 pairs (2 pairs) of edge states cross the gap at zero ($\pm\pi/T$) quasienergy.} 
    \label{fig:ribbon}
\end{figure}

In Fig.~\ref{fig:phase} we show the two topological numbers $C_0$ and $C_{\pi}$ as functions of $T\gamma$ and $\alpha = eA_0a_0/c$, where $A_0$ is the amplitude of the drive and $a_0$ is the lattice constant. An intricate ``phase diagram'' of topological transitions is observed. In a finite system with an edge, the number of chiral edge state at quasienergy $\varepsilon$ is equal to $|C_{\varepsilon}|$. In Fig.~\ref{fig:ribbon} we show the appearance of respectively $|C_0|$ and $|C_{\pi}|$ chiral edge states in a ribbon geometry for a point in the phase diagram of Fig.~\ref{fig:phase}. We have checked that the edge states are separated along the two opposite edges according to their chirality.

%--------- Fig S1 ------------%
%-----------------------------%

%--------- Fig S2 ------------%
%-----------------------------%

\subsection{Perturbation theory at $\Gamma$ point}

Here, we report the results of our perturbation theory near the first $\Gamma$ point transition at the FZ boundary where $\Omega/2 = 3\gamma J_0(\alpha)$. To second order in ${\bf k}$, the projected
Hamiltonian reads $\overline{H}_F={\Omega \over 2}+{\bf h}_1({\bf k})\cdot\boldsymbol{\sigma}$, where
%%%%%
\begin{eqnarray}
h_{x}-ih_{y} &=& -i{{3\gamma} \over 8} J_1(\alpha)(k_x-ik_y)^2a_0^2, \\
h_{z} &=& -{\Omega \over 2}+3\gamma J_0(\alpha)(1-\frac14k^2a_0^2),
\end{eqnarray}
%%%%%
The off-diagonal $(k_x\pm ik_y)^2$ term results in a change of Chern number $|\Delta C| = 2$ across the gap closing transition.
This is precisely the change we observe in our numerical calculations around $T\gamma\approx2.1$ for $\alpha=1.5$.

\subsection{Floquet Green's function method}
Here we sketch the details for the derivation of the conductance and numerical computations described in the main text. We employ the Green's function approach~\cite{arrachea2002,arrachea2005,arrachea2006,Sam73a} for deriving the charge current in the system described by time dependent Hamiltonian $H_{\mathrm{w}}(t)$ and contact Hamiltonian $H_{\mathrm{c}}$,
\begin{align}
& H_{\mathrm{w}}(t) = \sum_{l,m} A_{l m}(t)c^{\dagger}_lc_m \equiv c^{\dagger} A(t) c\nonumber \\
& H_{\mathrm{c}}  = \sum_{\lambda,\alpha}K^{\lambda}_{\alpha l}a^{\lambda \dagger}_{\alpha}c_l + K^{\lambda *}_{\alpha l} c^{\dagger}_l a^{\lambda}_{\alpha} \equiv \sum_{\lambda}a^{\lambda\dagger}K^{\lambda}c +\text{h.c.},\nonumber
\end{align}
where $c^{\dagger}_l$ and $a^{\lambda\dagger}_{\alpha}$ respectively denote the creation operator for electron at site $l$ of the system and site $\alpha$ of the lead $\lambda$. The net charge current flowing across the contact $\lambda$ into the wire is ($\hbar=1$)
\begin{align}
 J^{\lambda} (t) &= i e\left[H_{\mathrm{w}}(t)+H_{\mathrm{c}},N^{\lambda}(t)\right] \nonumber \\
 &= i e\left( c^{\dagger}(t)K^{\lambda\dagger}a^{\lambda}(t) - \mathrm{h.c.} \right),\label{spl:curr-def}
\end{align}
where $N^\lambda$ is the number operator for electrons in lead $\lambda$. Solving the Heisenberg equation for  $a^{\lambda} (t)$, we have
\begin{align}\label{spl:at}
 a^{\lambda}(t) = \eta^{\lambda}(t) + \int_{t_0 \rightarrow -\infty}^t g^{\lambda} (t-t') K^{\lambda} c (t') d t',
\end{align}
where, $t_0$ is the \textit{switching time} and $g^{\lambda} (t-t')$ is the Green's function matrix in lead $\lambda$. The \textit{noise} term  $\eta^{\lambda} (t) = i g^{\lambda} (t-t_0)a^{\lambda} (t_0)$ obeys the fluctuation-dissipation relation after averaging over the lead states
\begin{align}
 &\langle \eta^{\lambda \dagger}_l(\omega)\eta^{\lambda'}_{l'}(\omega') \rangle = (2\pi)^2 \delta_{\lambda \lambda'}\rho^{\lambda}_{ll'}(\omega)f_{\lambda}(\omega)\delta(\omega - \omega'), \\
 &\langle \eta^{\lambda}_l(\omega)\eta^{\lambda'\dagger}_{l'}(\omega') \rangle = (2\pi)^2 \delta_{\lambda \lambda'}\rho^{\lambda}_{ll'}(\omega)\bar f_{\lambda}(\omega)\delta(\omega - \omega'),
\end{align}
where $\rho^{\lambda}(\omega) = - \frac{1}{\pi}\textrm{Im}[g^{\lambda}(\omega)]$ is the density of states at lead $\lambda$ and  $f_{\lambda}(\omega)= 1-\bar f_\lambda(\omega) = \left[1+e^{(\omega-eV_{\lambda})/\tau_{\lambda}}\right]^{-1}$ is the Fermi-Dirac distribution of the  lead $\lambda$ with bias $V_{\lambda}$ and temperature $\tau_{\lambda}$. (The Boltzmann constant $k_B=1$.)

Integrating the Heisenberg equation for the electronic operator in the driven system gives~\cite{kohler2005}
\begin{align}
 \left[i \frac{\partial}{\partial t}  - i A(t) \right]c(t) - i \int_0^{\infty}d s~\Gamma(s)c(t-s) = h(t),
\end{align}
with self energy  $i\Gamma (s) =\sum_{\lambda} K^{\lambda \dagger} g^{\lambda}(s) K^{\lambda}$ and $h(t)  =\sum_{\lambda} K^{\lambda \dagger} \eta^{\lambda}(t)$. 
The Green's function $G(t,t')$  of this inhomogeneous equation satisfies
\begin{align}\label{eq:GFF}
 \left[i \frac{\partial}{\partial t}  - i A(t) \right]G(t,t') - &i \int_0^{\infty}d s~\Gamma(s)G(t-s,t')d s \nonumber \\
 &= \delta(t-t').
\end{align}
For a periodic drive with period $T = 2\pi/\Omega$, the \textit{Floquet Green's function} is also periodic over the same period $G(t+T,t'+T) = G (t,t')$. One can introduce the Fourier transform,
\begin{align}
G(t,t') &= \sum_{k\in\mathbb{Z}}\int \frac{d\omega}{2\pi}~ G^{(k)}(\omega) e^{-i\omega(t-t')}e^{-i k\Omega t},
\end{align}
The electronic operator in the system $c(t)$ is solved in terms of the Green's function
\begin{align}
 c(t) = \sum_k \int \frac{d\omega}{2\pi} e^{-i\omega t}e^{-i k\Omega t}G^{(k)}(\omega)h(\omega).
\end{align}
For a two terminal system ($\lambda = L,R$), using these expressions in the current formula  Eq.~(\ref{spl:curr-def}), the steady state current $I \equiv \int_0^T \left<{J^L(t)}\right> d t/T$ is found to be
\begin{align}
I = \frac{e}{2\pi }\int d\omega\sum_k\left[ T^{(k)}_{LR}(\omega)f_{R}(\omega) - T^{(k)}_{RL}(\omega)f_{L}(\omega)\right],
\end{align}
with
\begin{align}
 T^{(k)}_{\lambda \lambda'}(\omega) &= \text{Tr}\left[G^{(k)\dagger}(\omega)\xi^{\lambda}(\omega+k\Omega)G^{(k)}(\omega)\xi^{\lambda'}(\omega)  \right], \nonumber
\end{align}
where $\xi^{\lambda}(\omega) =2\pi K^{\lambda \dagger}\rho^{\lambda}(\omega)K^{\lambda}$.
If we bias the leads symmetrically with chemical potentials in the left and right leads respectively $eV/2$ and $-eV/2$, then the differential conductance at zero temperature is
\begin{align}\label{spl:condF}
 \frac{dI}{dV} = \frac{e^2}{2\pi}\sum_k \left[ T^{(k)}_{LR}(-V/2) + T^{(k)}_{RL}(V/2)\right].
\end{align}

In our numerics, we model the leads and the system in the same graphene sheet by doping the leads at 1/6 of the band-width. 
We solve Eq.~(\ref{eq:GFF}) in the ``wide band limit'' when the density of states $\rho^{\lambda}$ of lead $\lambda$ is constant for the relevant energy scales, in which case $\Gamma(s)\propto\delta(s)$. We compute the Green's function of the lead $g^{\lambda}$ through the recursive Green's function method~\cite{Sancho}. 
Along with this we have broadened the bands to have the density of states remain effectively the same for a wide range of energy, thus ensuring the wide band limit is applicable. We have checked that the qualitative features of our results do not depend on the details of modeling the lead.

\end{document}